\documentclass[journal=jacsat,manuscript=communication,etalmode=truncate,keywords=true]{achemso}
\setkeys{acs}{articletitle=true,maxauthors=10,etalmode=truncate,keywords=true}
\usepackage{amsmath,amssymb,units,txfonts}
\usepackage{amsfonts,times}
\usepackage{graphicx,multirow}
\usepackage[usenames,dvipsnames]{xcolor}
\definecolor{StyleColor}{RGB}{34,80,169} 
\definecolor{abstractcolor}{RGB}{255,243,201}
\definecolor{DarkGreen}{RGB}{0,100,5}
\usepackage[bookmarks=true,colorlinks=true,urlcolor=StyleColor,linkcolor=StyleColor,citecolor=StyleColor]{hyperref} 
\usepackage{colortbl,cuted}
\newcommand{\tripletoxygen}{C$\cdots$O$_2^t$$\cdots$C}
\newcommand{\ketene}{C$=$C$=$O}
\newcommand{\singletoxygen}{C$-$O$^s_2$$-$C}
\newcommand{\carbonyl}{C$_2$$>$C$=$O}

\newcommand{\epoxyh}{C$_2^{h}$$>$O}
\newcommand{\epoxyp}{C$_2^{p}$$>$O}
\newcommand{\epoxy}{C$_2$$>$O}

\makeatletter\newenvironment{abstractbox}{
   \begin{lrbox}{\@tempboxa}\begin{minipage}{\textwidth}}{\end{minipage}\end{lrbox}
   \colorbox{abstractcolor}{\usebox{\@tempboxa}}
}\makeatother

\renewcommand*\authorfont{\flushleft}
\renewcommand*\affilfont{\mdseries\flushleft}
\renewcommand*\titlesize{\Large}
\renewcommand*\titlefont{\flushleft\sf\bfseries}
\def\refname{\sffamily\color{StyleColor}\Large$\blacksquare$\normalsize\ REFERENCES}
\usepackage{titlesec}
\titleformat{\section}{\bfseries\sffamily\color{StyleColor}}{\thesection.~}{0pt}{}
\titleformat{\subsection}[runin]{\bfseries\sffamily\normalsize}{\indent\thesubsection.~}{0pt}{}[.]
\titlespacing{\subsection}{0pt}{0pt}{*1}
\titleformat{\subsubsection}{\bfseries\sffamily\normalsize}{\thethesubsection.~}{0pt}{}
\titlespacing{\subsubsection}{0pt}{0pt}{*0}

\usepackage{amsmath}
\usepackage{amssymb}
\usepackage{wasysym}
\usepackage{graphicx}
\usepackage{cancel}
\usepackage{xcolor}
\usepackage[latin1]{inputenc}
\usepackage{times}

\title{Disentangling Vacancy Oxidation on Metallicity-Sorted Carbon Nanotubes}

\author{Duncan J. Mowbray}
\affiliation[]{\footnotemark[2]{\ }
Nano-Bio Spectroscopy Group and ETSF Scientific Development Centre, Departamento de F{\'{\i}}sica de Materiales, Centro de F{\'{\i}}sica de Materiales CSIC-UPV/EHU-MPC and DIPC, Universidad del Pa{\'{\i}}s Vasco UPV/EHU, E-20018 San Sebasti\'{a}n, Spain}
\email{duncan.mowbray@gmail.com}
\author{Alejandro P\'{e}rez Paz}
\affiliation[]{\footnotemark[2]{\ }
Nano-Bio Spectroscopy Group and ETSF Scientific Development Centre, Departamento de F{\'{\i}}sica de Materiales, Centro de F{\'{\i}}sica de Materiales CSIC-UPV/EHU-MPC and DIPC, Universidad del Pa{\'{\i}}s Vasco UPV/EHU, E-20018 San Sebasti\'{a}n, Spain}
\email{alejandroperezpaz@yahoo.com}
\author{Georgina Ruiz-Soria}
\affiliation[]{\newline\footnotemark[3]{\ } Faculty of Physics, University of Vienna,
Strudlhofgasse 4, A-1090 Vienna, Austria}
\author{Markus Sauer}
\affiliation[]{\newline\footnotemark[3]{\ } Faculty of Physics, University of Vienna,
Strudlhofgasse 4, A-1090 Vienna, Austria}
\author{Paolo Lacovig}
\affiliation[]{\newline\footnotemark[5]{\ } Elettra Sincrotrone Trieste, S.S.~14 km 163.5, 34149 Trieste,
Italy}
\author{Matteo Dalmiglio}
\affiliation[]{\newline\footnotemark[5]{\ } Elettra Sincrotrone Trieste, S.S.~14 km 163.5, 34149 Trieste,
Italy}
\author{Silvano Lizzit}
\affiliation[]{\newline\footnotemark[5]{\ } Elettra Sincrotrone Trieste, S.S.~14 km 163.5, 34149 Trieste,
Italy}
\author{Kazuhiro Yanagi}
\affiliation[]{\newline\footnotemark[4]{\ } Department of Physics, Tokyo Metropolitan University,
Hachioji, 192-0397 Tokyo, Japan}
\author{Andrea Goldoni}
\affiliation[]{\newline\footnotemark[5]{\ } Elettra Sincrotrone Trieste, S.S.~14 km 163.5, 34149 Trieste,
Italy}
\author{Thomas Pichler}
\affiliation[]{\newline\footnotemark[3]{\ } Faculty of
Physics, University of Vienna, Strudlhofgasse 4, A-1090 Vienna,
Austria}
\author{Paola Ayala}
\affiliation[]{\newline\footnotemark[3]{\ } Faculty of Physics, University of Vienna,
Strudlhofgasse 4, A-1090 Vienna, Austria}
\alsoaffiliation[]{\newline\footnotemark[6]{\ } School of Physical Sciences and Nanotechnology, Yachay Tech University, Urcuqu\'{\i} 100119, Ecuador}
\author{Angel Rubio}
\affiliation[]{\footnotemark[2]{\ }
Nano-Bio Spectroscopy Group and ETSF Scientific Development Centre, Departamento de F{\'{\i}}sica de Materiales, Centro de F{\'{\i}}sica de Materiales CSIC-UPV/EHU-MPC and DIPC, Universidad del Pa{\'{\i}}s Vasco UPV/EHU, E-20018 San Sebasti\'{a}n, Spain}
\alsoaffiliation[MaxPlanck]{\newline$^\perp$Max Planck Institute for the Structure and Dynamics of Matter and Center for Free-Electron Laser Science \& Department of Physics, University of Hamburg, Luruper Chaussee 149, D-22761 Hamburg, Germany}
\email{angel.rubio@ehu.es}

\begin{document}
\maketitle
\begin{strip}
\vspace{-1cm}

\noindent{\color{StyleColor}{\rule{\textwidth}{0.5pt}}}
\begin{abstractbox}
\begin{tabular*}{17cm}{b{11.5cm}r}
\noindent\textbf{\color{StyleColor}{ABSTRACT:}}
Pristine single-walled carbon nanotubes (SWCNTs) are rather inert to O$_2$ and N$_2$, which for low doses chemisorb only on defect sites or vacancies of the SWCNTs at the ppm level.  However, very low doping has a major effect on the electronic properties and conductivity of the SWCNTs.  Already at low O$_2$ doses (80 L), the X-ray photoelectron spectroscopy (XPS) O 1s signal becomes saturated, indicating nearly all the SWCNT's vacancies have been oxidized.   As a result, probing vacancy oxidation on SWCNTs via XPS yields spectra with rather low signal-to-noise ratios, even for metallicity-sorted SWCNTs. We show that, even under these conditions, the first principles density functional theory calculated Kohn-Sham O 1s binding energies may be used to assign the XPS O 1s spectra for oxidized vacancies on SWCNTs into its individual components.  This allows one to determine the specific functional groups or bonding environments measured.  We find the XPS O 1s signal is mostly due to three O-containing functional groups on SWCNT vacancies: epoxy (\epoxy), carbonyl (\carbonyl), and ketene~(\ketene), as ordered by abundance.  Upon oxidation of nearly all the SWCNT's vacancies, the central peak's intensity for the metallic 
&\includegraphics[width=2in]{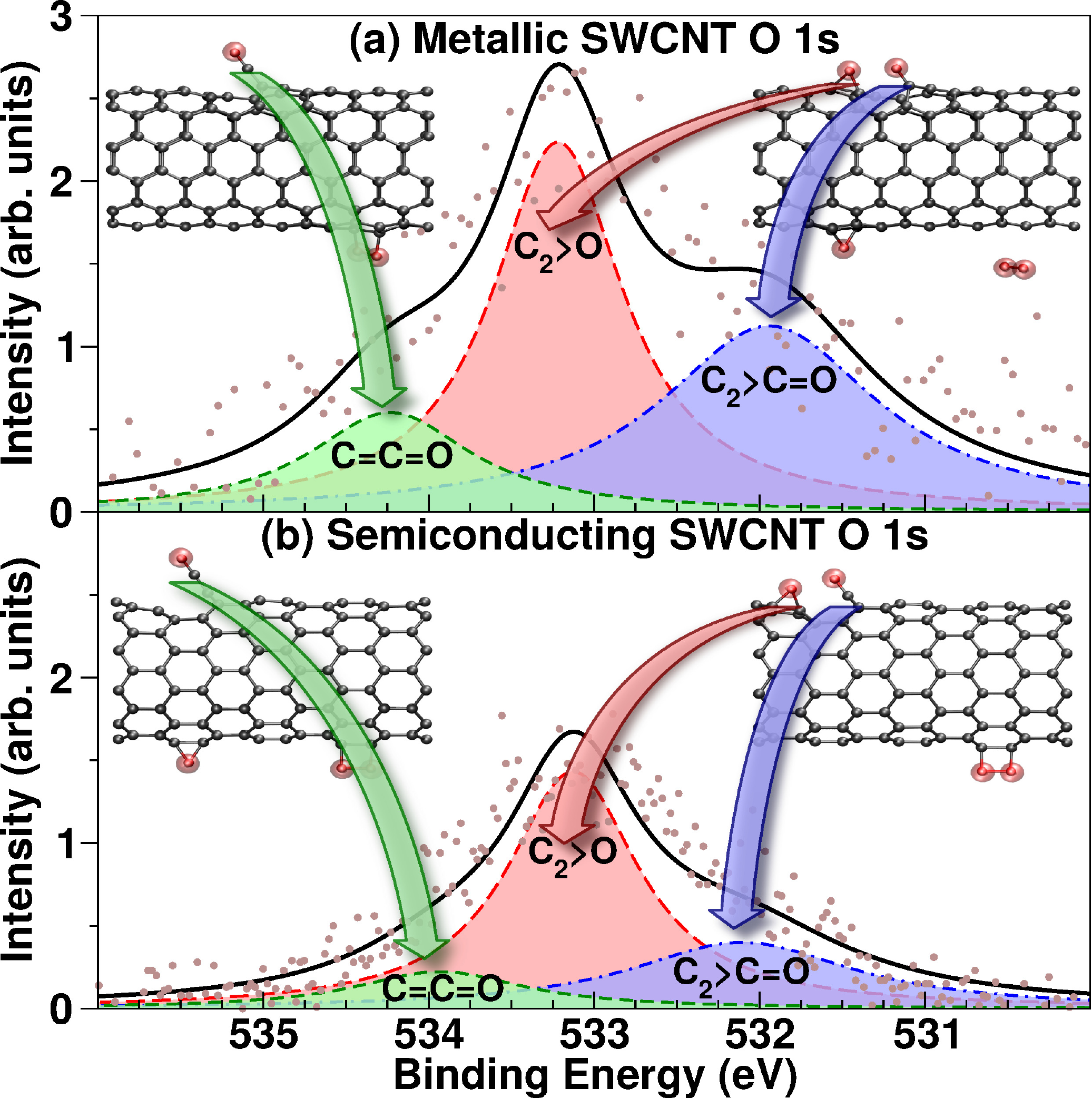}\\
\multicolumn{2}{p{17cm}}{
SWCNT    sample is 60\% greater than for the semiconducting SWCNT sample.  This suggests a greater abundance of O-containing defect structures on the metallic SWCNT sample.  For both metallic and semiconducting SWCNTs, we find O$_2$ does not contribute to the measured XPS O~1s spectra.  
}
\end{tabular*}
\end{abstractbox}
\noindent{\color{StyleColor}{\rule{1.\textwidth}{0.5pt}}}
\end{strip}

\section{INTRODUCTION}
Since their discovery in the early 1990s\cite{Iijima}, single-walled carbon nanotubes (SWCNTs) have been promising materials for electronic applications due to their ballistic transport, long coherence length, and high thermal conductivity in vacuum\cite{Harris, Dresselhaus,JuanmaPRL}.  Although pristine SWCNTs are rather inert to O$_2$ and N$_2$,\cite{TomaOdissociationSWNT,OdissociationSWNT} exposure to air oxidizes the SWCNT's vacancies, as was recently shown for monovacancies in graphene\cite{O2dissociationGrapheneMV}.  More importantly, vacancy oxidation leads to significant changes in both impurity scattering and the overall conductivity of the SWCNT\cite{GarciaLastra10PRB}.  This property can be successfully exploited to construct SWCNT-based gas detecting nanodevices \cite{Bondavalli09SABC,Jhi00PRL,Collins00S,Ulbricht02PRB,Goldoni03JACS}.   Still, the precise structure and composition of SWCNTs after O$_2$ exposure remains the subject of much debate\cite{Jhi00PRL,Collins00S,Ulbricht02PRB,Dag2003,Grujicic03ASS,Giannozzi03JCP,Goldoni03JACS,Datsyuk08C,Mowbray09PRB,GarciaLastra10PRB,WandaAndreoni,ACSNano}.  

From the experimental viewpoint, this is in part due to impurities, mixed metallicity samples, sample inhomogeneities, and differences between the experimental protocols and physical/chemical conditions used.  As a result, discrepancies exist in the literature regarding the impact of vacancy oxidation on the electronic, transport and optical properties of carbon allotropes.  For example, it is still an open question as to how the O~1s X-ray photoelectron spectroscopy (XPS) spectra of carbon allotropes exposed to O$_2$ should be disentangled. 

From the computational viewpoint, the size and time scales needed to model the process of vacancy oxidation, the role of the environment, and the sample's quality require a multiscale approach combining first principles density functional theory (DFT) and semiempirical coarse grained methods.  We will sidestep these problems in the present work by instead focusing on correlating the measured O~1s XPS signal to the different species potentially formed on the sample.  However, even sophisticated approaches, such as final state methods, often yield core level energies with absolute errors larger than the XPS measured shifts.  Nevertheless, the calculated core level shifts agree quite well with XPS measurements.  Although this sometimes requires a nonlocal description of exchange and correlation \cite{HybridCLSJPC2014,HybridCLSPRB2012}, this is not universally the case \cite{PBEvsPBE0CLSJPCC2012,PBEvsPBEUJPCC2015,Cabellos12JPCC}.  Specifically, we have previously shown that a semilocal description of exchange and correlation is sufficient to describe C 1s core level shifts in vacancy oxidized SWCNTs \cite{ACSNano}.  However, this necessitates the use of an absolute experimental energy reference to align the calculated peaks.  Only through this interplay between theory and experiment can one obtain an absolute assignment of specific O species on SWCNTs.   

However, to obtain a high signal-to-noise ratio, previous XPS O~1s studies for carbonaceous compounds have focused on the high coverage regime\cite{kinga, rozada, Vinogradov, Datsyuk08C, LangmuirO1s2005,Martinez03,Markiewicz_PCCP2014}, where the sample is heavily oxidized. Table~\ref{Table0}
\begin{table}[!t]
\caption{\rm{\bf{Measured XPS O~1s Binding Energies $\boldsymbol\varepsilon_\text{O1s}$ in Electronvolts of Oxidized Graphite,  Graphene, and Multi-Walled (MW) and Single-Walled (SW) Carbon Nanotubes (CNTs)}}}\label{Table0}
\begin{tabular}{lccc}
\multicolumn{4}{>{\columncolor[gray]{0.9}}c}{ }\\[-3mm]
\multicolumn{1}{>{\columncolor[gray]{0.9}}c}{material} & 
\multicolumn{1}{>{\columncolor[gray]{0.9}}c}{assignment} & 
\multicolumn{1}{>{\columncolor[gray]{0.9}}c}{O bond type} & 
\multicolumn{1}{>{\columncolor[gray]{0.9}}c}{$\varepsilon_\mathrm{O1s}$ (eV)}\\
graphite oxide & carbonyl & double & 531.9$^a$\\
graphite oxide$^b$ & carbonyl & double & 531.7$^a$\\
graphite oxide & hydroxyl/epoxy & single & 533.1$^a$\\
graphite oxide$^b$ & hydroxyl/epoxy & single & 533.5$^a$\\
graphite oxide$^b$ & carbonyl/carboxyl & double & 530.5$^{c}$\\
graphite oxide$^b$ & hydroxyl/epoxy & single & 531.8$^{c}$\\
graphite oxide$^b$ & ester/carboxyl & single & 533.3$^{c}$\\
graphene & epoxy & single & 531.1$^d$\\
graphene$^e$ & carbonyl/carboxyl & double & 532.7$^d$\\
MWCNT & carbonyl/carboxyl & double& 531.9$^{f}$\\
MWCNT & hydroxyl/epoxy & single & 533.2$^{f}$\\
MWCNT & carbonyl/carboxyl &double & 531.3$^g$\\
MWCNT & hydroxyl/epoxy &single & 533.0$^g$\\
SWCNT & carbonyl/carboxyl & double & 531.6$^h$\\
SWCNT & hydroxyl & single & 533.3$^h$\\
SWCNT & carbonyl/carboxyl & double & 532.04$^i$\\
SWCNT & hydroxyl/epoxy & single & 533.41$^i$\\
\multicolumn{4}{p{0.95\columnwidth}}{\footnotesize$^a$Reference~\citenum{kinga}.  $^b$Thermally reduced.  $^c$Reference~\citenum{rozada}. $^d$Reference~\citenum{Vinogradov}.  $^e$Sputtered with Ar$^+$.  $^f$Reference~\citenum{Datsyuk08C}.  $^g$Reference~\citenum{LangmuirO1s2005}. $^h$Reference~\citenum{Martinez03}\nocite{Martinez03}. $^i$Reference~\citenum{Markiewicz_PCCP2014}\nocite{Markiewicz_PCCP2014}.}
\end{tabular}
\noindent\color{StyleColor}{\rule{\columnwidth}{1.0pt}}
\end{table}
 provides a brief survey of experimental XPS O~1s binding energies in oxidized carbonaceous compounds: graphite oxide\cite{kinga,rozada}, graphene\cite{Vinogradov}, multi-walled (MW) CNTs\cite{Datsyuk08C,LangmuirO1s2005}, and SWCNTs\cite{Martinez03,Markiewicz_PCCP2014}.  The O~1s spectra are typically decomposed into components associated with singly and doubly bonded O atoms.   However, the precise structure of the O-containing functional groups remains unclear.  

Singly bonded O atoms may be present as hydroxyl, epoxy, ester, or carboxyl groups, while doubly bonded O atoms may be present as carbonyl, ketene, ester, or carboxyl groups.  Furthermore, carbonates may be present as contaminants, complicating  the XPS O~1s spectrum's decomposition.  For example, the assignment in ref~\citenum{rozada} of a peak at 530.5~eV to carbonyl is more likely the result of carbonates\cite{Datsyuk08C}, with the remaining peaks at 531.8 and 533.3~eV associated with carbonyl/carboxyl and hydroxyl/epoxy, respectively.  Moreover, the assignment for graphene on metal surfaces\cite{Vinogradov} is opposite to that found for graphite oxide and oxidized CNTs in the literature.  Taking this into account, we obtain for graphite oxide and CNTs consistent O~1s binding energies of 531.3 to 532.04~eV for doubly bonded O and 533.1 to 533.5~eV for singly bonded O.   It is clear from Table~\ref{Table0} that the assignment of O~1s spectra for carbon-based materials remains a controversial issue given the high variability of the values reported.

In this article, we use first principles simulations together with XPS experiments to tackle this controversial topic.  Specifically, we disentangle the O~1s XPS spectra of metallicity-sorted SWCNTs for O$_2$ exposures, including both monovacancies and the pristine surface.   In this way, we are able to identify the most common species present on SWCNTs upon vacancy oxidation.  This article is organized as follows.  In Section \ref{methods}  we describe the experimental setup, computational procedure, and theoretical models we employed to measure, model, and fit the XPS spectra.  The resulting decomposed spectra are discussed and analyzed in Section \ref{resultsanddiscussion}.  This is followed by concluding remarks in Section \ref{conclusions}.

\section{METHODS}\label{methods}
\subsection{Experimental Details} The experiments described herein were performed according to the same protocol and on the same samples as those described in ref~\citenum{ACSNano}.  These SWCNTs were synthesized by the
arc-discharge method, purified and separated into metallic and
semiconducting tubes, and then deposited on sapphire substrates. XPS 
studies were conducted at the SuperESCA beamline at the
ELETTRA synchrotron~\cite{Abrami95RSI}, where the emitted
photoelectrons are collected by a 150~mm hemispherical analyzer
with a time-delay detector mounted at 70$^\circ$ with respect to the
incident beam\cite{ACSNano}. The samples were placed on a vertical Ta holder,
mounted on a manipulator that allows cooling down to about 100 K
and annealing up to 1800 K. The experimental chamber had a base
pressure of $2.5\times10^{-10}$~mbar. In order to remove any
remaining impurity, the samples were outgassed in-situ by a
combined resistive and electron-beam heating system up to 900 K.
The exceptional purity of the samples was confirmed by wide range
high resolution (HR) photoemission spectroscopy (PES) survey
scans~\cite{Ayala09PRB,Kramberger07PRB}.  The
different nanotube samples were exposed to pure O$_2$ gas, 
which was inserted through a needle valve.  
Core level spectra were recorded for the O~1s signal using a photon energies of 650~eV,
 with corresponding overall energy resolution of 200~meV. Ta was used for calibration
measurements. 

During the gas sensing experiments we kept the temperature of the samples constant at 100 K, and the gas pressure below 10$^{-8}$ mbar. Gas doses of 60 and 80 L (1L $\approx$ 1.33$\times$10$^{-6}$ mbar$\cdot$s) were used. These very low oxygen doses allow us to preferentially probe vacancy oxidation within the SWCNT samples.  Because the O~1s spectra were unchanged between these two doses, this demonstrates that defects are already saturated at a dose of 80 L, which we will focus on from here on.

\subsection{Computational Details}\label{Computational Details}
DFT calculations were performed using the projector augmented wave function (PAW) method implemented in the \textsc{GPAW} code \cite{Mortensen05PRB,Enkovaara10JPCM,Bahn02CSE} within the local density approximation (LDA) \cite{Perdew81PRB} for the exchange and correlation (xc)-functional. We employed a grid spacing of $h\approx 0.2$ \AA, a \textbf{k}-point sampling of ($1\times1\times3$), and an electronic temperature of 0.1~eV with all energies extrapolated to $T\rightarrow0$~K, that is, neglecting the electronic entropy term in the calculated free energy $-ST$.  All
structures were relaxed until a maximum force below 0.05~eV/\AA{} was reached. We included spin-polarization whenever necessary, for example, for open-shell systems.   

All-electron PAW calculations were used to obtain O~1s binding energies within the initial state method\cite{Cabellos12JPCC,ACSNano}, that is, from the ground state Kohn-Sham (KS) O~1s eigenvalues.  This method neglects electronic relaxation effects in the presence of the hole that are included in the final state method\cite{Cabellos12JPCC,TomaPRB2015}. To assess the impact of these effects on the O~1s binding energies in SWCNTs, we also employed the recently developed all-electron delta self-consistent field ($\Delta$SCF) final state method of ref~\citenum{TomaPRB2015}, employing the revPBE\cite{revPBE} xc-functional on our LDA structures.  This all-electron $\Delta$SCF final state method has been shown to provide absolute C~1s binding energies in quantitative agreement with those measured with XPS for graphene\cite{TomaPRB2015}.

A constant shift of 29.57~eV has been applied to the KS O~1s eigenvalues for the (6,6) and (10,0) SWCNTs.  This was chosen to align the calculated carbonyl O~1s peaks with their experimental counterparts at 531.9~eV reported in refs~\citenum{kinga}, \citenum{rozada}, and \citenum{Datsyuk08C}. 
We followed this approach rather than attempting to match the experimental O~1s value for O$_2$~\cite{Sorensen} due to the large renormalization of the molecular O~1s after physisorption.   Note that the applied shift, although large in absolute terms, amounts to a relative error of 5.5\% in the absolute O~1s binding energy, which is consistent with previous initial state binding energy shifts \cite{Cabellos12JPCC,ACSNano}.  As we will show in Section~\ref{initialvsfinalstatemethods}, an all-electron $\Delta$SCF final state calculation provides more accurate absolute O~1s binding energies, but still requires a shift to obtain the absolute alignment. 

We consider as prototype defective semiconducting and metallic nanotubes the (10,0)
and (6,6) SWCNTs with 0.58 monovacancies per nanometer.  Considering the localized nature of the O~1s, the conclusions drawn from these highly defective small diameter SWCNTs ($d\approx 8$~\AA) should also be applicable to larger diameter experimentally relevant SWCNTs with fewer defects. A vacuum layer of more
than 10\ \AA\ was used to remove interactions between repeated images
perpendicular to the tube axis.  The (10,0) and (6,6) SWCNTs with a
monovacancy were modeled with 17.052 and 17.229~\AA\ long cells, respectively.  
To ensure an absolute energy reference for the O~1s core-levels between calculations, nonperiodic
boundary conditions have been employed perpendicular to the nanotube axis by fixing the electron density and KS wave functions to zero at the cell boundary.

From the computational point of view, the O~1s levels are strongly localized on the O atom, with little interaction beyond next-nearest neighbor.  This means we can obtain the O~1s KS eigenenergies for  multiple O species, separated by more than 8 C atoms, within the same calculation.  As a further check, we included a chemisorbed O$_2$ molecule in several calculations, and found that its O~1s binding energies differ by less than 50~meV.   The charges of the various O species were computed by applying a Bader analysis\cite{Henkelman2006354} to the all-electron charge density.

\subsection{Fitting Procedure}\label{FittingProcedure}

We use a sum of Voigtian functions $V(I, \Gamma, \sigma;\varepsilon)$  to fit the measured XPS O~1s spectra.  These Voigtians are the convolution of a Gaussian describing the resolution of the experimental apparatus $\sigma$, and a Lorentzian describing the inverse lifetime of the core-hole excitation $\Gamma$ centered on the DFT O~1s binding energies $\varepsilon_{\mathrm{O1s}}$.
  Specifically,
\begin{equation}
V(I,\Gamma,\sigma;\varepsilon) = I\int_{-\infty}^\infty d\varepsilon' \frac{e^{-\frac{\varepsilon'^2}{2\sigma^2}}}{\sqrt{2\pi}\sigma} \frac{(\Gamma/2)^2}{(\varepsilon-\varepsilon')^2 + (\Gamma/2)^2},\label{voigtiandef} 
\end{equation}
where $I$ is the intensity of the Lorentzian and $\Gamma$ is the inverse lifetime, that is, the full width at half maximum (FWHM), and $\varepsilon$ is the binding energy relative to $\varepsilon_{\mathrm{O1s}}$ for a particular O species.  The experimental resolution of the apparatus, $\sigma$, yields an effective intensity of the individual contributions to the O~1s XPS spectra of 
\begin{equation}
I^{\mathit{eff}} = V(I,\Gamma,\sigma;\varepsilon=0)=I\sqrt{\pi} \alpha e^{\alpha^2} \mathrm{erfc}\left(\alpha\right),
\label{Ieff}
\end{equation}
where $\alpha=\frac{\Gamma/2}{\sqrt{2}\sigma}$ is a dimensionless parameter and $\mathrm{erfc}\left(\alpha\right)$ is the complementary error function. 
As the experimental resolution increases, that is, $\sigma$ decreases, the effective intensity approaches the Lorentzian intensity quadratically from below, 
\begin{equation}
I^{\mathit{eff}}\approx I\left[1-\frac{\sigma^2}{\left(\Gamma/2\right)^2}+\mathcal{O}\left(\sigma^4\right)\right].
\end{equation}
In our experimental setup $\sigma\approx50$~meV in the O~1s region. For typical inverse lifetimes of an electronvolt, $I$ and $I^{\mathit{eff}}$ differ by less than 1\%.  Using the correlation coefficient for the Voigtian fit as a guide, we can identify which species are most abundant on the SWCNTs.  In this way, we unravel the XPS O~1s spectra into individual O-containing species on the SWCNTs. 

\section{RESULTS AND DISCUSSION}\label{resultsanddiscussion}

\subsection{Possible Oxygen Species on SWCNTs}

The first task in our theoretical disentanglement of the O~1s XPS spectra is building a database of O~1s binding energies for potential O-containing species or functional groups on the SWCNTs. Here, binding energies have been obtained by rigidly shifting the KS O~1s eigenenergies to align the calculated and measured carbonyl peaks \cite{kinga,rozada,Datsyuk08C,LangmuirO1s2005,PCCPO1sSWNTs2015}.  
  
\begin{figure}[!t]
\includegraphics[width=\columnwidth]{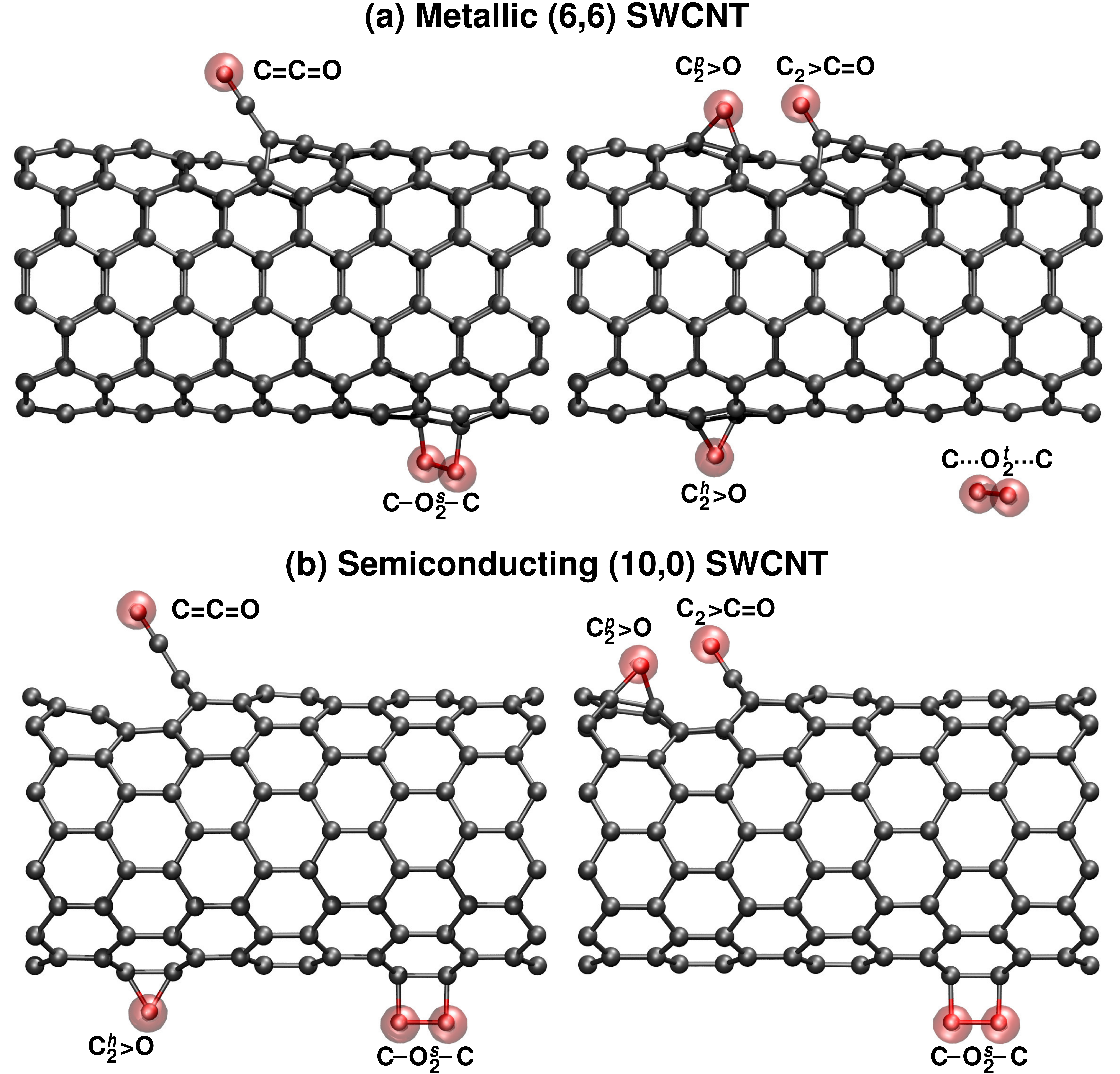}
\caption{Schematics of O-containing species on (a) metallic (6,6) and (b) semiconducting (10,0) SWCNTs: ketene (\ketene), epoxy on pentagon and hexagon sites (\epoxyp{} and \epoxyh), carbonyl (\carbonyl), and chemisorbed singlet and physisorbed triplet O$_2$ (\singletoxygen{} and \tripletoxygen). Isosurfaces (0.01~e/\AA$^{3/2}$ ) of the O~1s levels are shown in red, while C and O atoms are depicted as gray and red spheres, respectively.}
\label{Fig1}
\noindent{\color{StyleColor}{\rule{\columnwidth}{1.0pt}}}
\end{figure}

We consider physisorbed O$_2$ atop in the triplet state (\tripletoxygen), chemisorbed O$_2$ atop in the singlet state (\singletoxygen), and atomic O in an epoxy group on a hexagon (\epoxyh) away from the SWCNT's monovacancy, as shown in Figure~\ref{Fig1}.
Further, we consider ketene (\ketene), carbonyl (\carbonyl), and epoxy (\epoxyp) groups at the SWCNT's monovacancy, as shown in Figure~\ref{Fig1}.  The carbonyl and epoxy groups on the monovacancy are combined to model a dissociated O$_2$ molecule at a SWCNT's monovacancy.  This dissociation is known to be facile, with a barrier of 0.15~eV and exothermic by 4.95~eV on the graphene monovacancy\cite{O2dissociationGrapheneMV}.  In fact, we found adsorbing O$_2$ on the unoccupied (10,0) SWCNT's monovacancy induced a barrierless dissociation.  In this case, the epoxy group consists of an O atom bridging a side of a pentagon.  The ketene group on a monovacancy mimics an O atom that ``strips'' a C atom from the pristine SWCNT to form a monovacancy.

The calculated O~1s binding energies $\varepsilon_{\mathrm{O1s}}$, the charge of the O atom $Q_{\mathrm{O}}$, and the shortest O--C distance $d_{\mathrm{O-C}}$ for the O-containing species shown in Figure~\ref{Fig1} are given in Table~\ref{Table1}. 
\begin{table}
\caption{\rm{\bf{Calculated O~1s Binding Energies ${\boldsymbol\varepsilon}_{\text{O1s}}$ in Electronvolts, O Charge \textit{Q}$_{\text{O}}$ in \textit{e}, and O--C Separation \textit{d}$_{\text{O--C}}$ in \AA\ for Various O-Containing Species on Metallic (6,6) and Semiconducting (10,0) SWCNTs}}}\label{Table1}
\begin{tabular}{ccccc}
\multicolumn{5}{>{\columncolor[gray]{0.9}}c}{ }\\[-3mm]
\rowcolor[gray]{0.9}
symbol& species & $\varepsilon_\mathrm{O1s}$ & $Q_{\mathrm{O}}$ & $d_{\mathrm{O-C}}$\\
\rowcolor[gray]{0.9}
& & (eV) & ($e$) & (\AA)\\
\multicolumn{5}{c}{(6,6) SWCNT O~1s:}\\
\multirow{2}{*}{\tripletoxygen}    & physisorbed
&\multirow{2}{*}{535.58} 
&\multirow{2}{*}{$-0.03$} & 2.633\\
& triplet O$_2$ & & & 2.730\\
\ketene                    & ketene   
& 534.23 & $-0.73$ & 1.165\\
\multirow{2}{*}{\singletoxygen}              & chemisorbed
& \multirow{2}{*}{533.72}
& $-0.42$ & \multirow{2}{*}{1.465}\\
& singlet O$_2$ &  
& $-0.40$ & \\ 
\multirow{2}{*}{\epoxyh}         & epoxy on    
& \multirow{2}{*}{533.37} 
& \multirow{2}{*}{$-0.75$}
& \multirow{2}{*}{1.43} \\
& hexagon    & & \\
\multirow{2}{*}{\epoxyp}         & epoxy on 
&\multirow{2}{*}{533.22} 
&\multirow{2}{*}{$-0.73$} 
& 1.420\\
& pentagon &&& 1.436\\
\carbonyl                & carbonyl 
& 531.94 & $-0.82$ & 1.219\\
\multicolumn{5}{c}{(10,0) SWCNT O~1s:}\\
\ketene                    & ketene   
& 533.93 & $-0.73$ & 1.167\\
\multirow{2}{*}{\singletoxygen}              & chemisorbed
& \multirow{2}{*}{533.74} 
& $-0.48$ & 1.46\\
& singlet O$_2$ &  
& $-0.37$ & 1.46\\ 
\multirow{2}{*}{\epoxyh}         & epoxy on    
& \multirow{2}{*}{533.04}
& \multirow{2}{*}{$-0.73$}
& \multirow{2}{*}{1.44}\\
& hexagon    & \\
\multirow{2}{*}{\epoxyp}         & epoxy on 
& \multirow{2}{*}{533.13}
& \multirow{2}{*}{$-0.72$}& 1.442\\
& pentagon & & &1.433\\
\carbonyl                & carbonyl 
& 532.11 & $-0.81$ & 1.212\\
\end{tabular}
\noindent{\color{StyleColor}{\rule{\columnwidth}{1.0pt}}}
\end{table}
Core-level shifts are often explained in terms of charge transfer.  When charge is transferred into/out of an atom, the binding of the core-levels is reduced/increased.  Because O is an electronegative species, we find charge is transferred from the SWCNT to the O atoms, and the O~1s levels are always shifted to weaker binding relative to the nearly neutral physisorbed O$_2$ at 535.58~eV.  We also find the binding energy and net charge of chemisorbed singlet O$_2$ is nearly the same for both metallic (6,6) and semiconducting (10,0) SWCNTs (about $533.7$~eV and $-0.8e$).  On the other end, O in carbonyl has the weakest binding energy and the greatest charge on both SWCNTs of about $532.0$~eV and $-0.8e$.

However, the O~1s core-level shifts we observe cannot be entirely explained by this charge transfer argument alone.  For instance, for both metallic and semiconducting tubes the ketene and epoxy groups have similar charge transfers to the O atom of about $-0.7e$, while their binding energies differ by about 0.9~eV.  

To understand this behavior, one must also take into account the screening by the substrate, that is, the change in work function \cite{Cabellos12JPCC,MowbrayACSNano}.  This change in screening is related qualitatively to the difference in O--C bond length.  For the ketene group, where the O--C bond is 0.4~\AA\ shorter, the SWCNT is better able to screen the charge transfer to the O atom.  This results in a strengthening of the O~1s binding. 

Overall, charge transfer and screening shift the core-level binding energies in opposite directions, resulting in a partial cancellation of these effects.  As a result, it is often unclear what drives the measured core-level binding energies, as it is difficult to disentangle the effects of charge transfer and screening.  This necessitates a robust theoretical treatment to both partition these effects, and accurately describe the corresponding core-level shifts.  

\subsection{Comparison of Initial and Final State Methods}\label{initialvsfinalstatemethods}
To test the robustness of the initial state approximation, we have also calculated the O~1s binding energies using the screened all-electron core hole final state method \cite{TomaPRB2015}.  We obtained 530.85 and 529.58 eV for epoxy on a pentagon (\epoxyp) and carbonyl (\carbonyl), respectively, on the (6,6) SWCNT.  This is compared to our shifted DFT values of 533.22 and 531.94 eV in Table~\ref{Table1}.  In both cases, the energy differences are about 2.4 eV, yielding a relative error of 0.4\%. 

More importantly, we obtain the same core level shift in O~1s binding energy between carbonyl and epoxy on a pentagon from both methods.  This suggests that the final state method simply provides a renormalization of the absolute binding energy, while the physically relevant differences between structures are already accounted for by the initial state method.

Although the final state method improves the accuracy of the O~1s binding energy by an order of magnitude over the initial state method, it still requires a constant shift of 2.4~eV. At the same time, it has twice the computational cost of the initial state method.
This means the final state method provides no practical advantage over the initial state method for our systems.

\subsection{Fitting of O 1s XPS Spectra}

\begin{figure}[!t]
\includegraphics[width=\columnwidth]{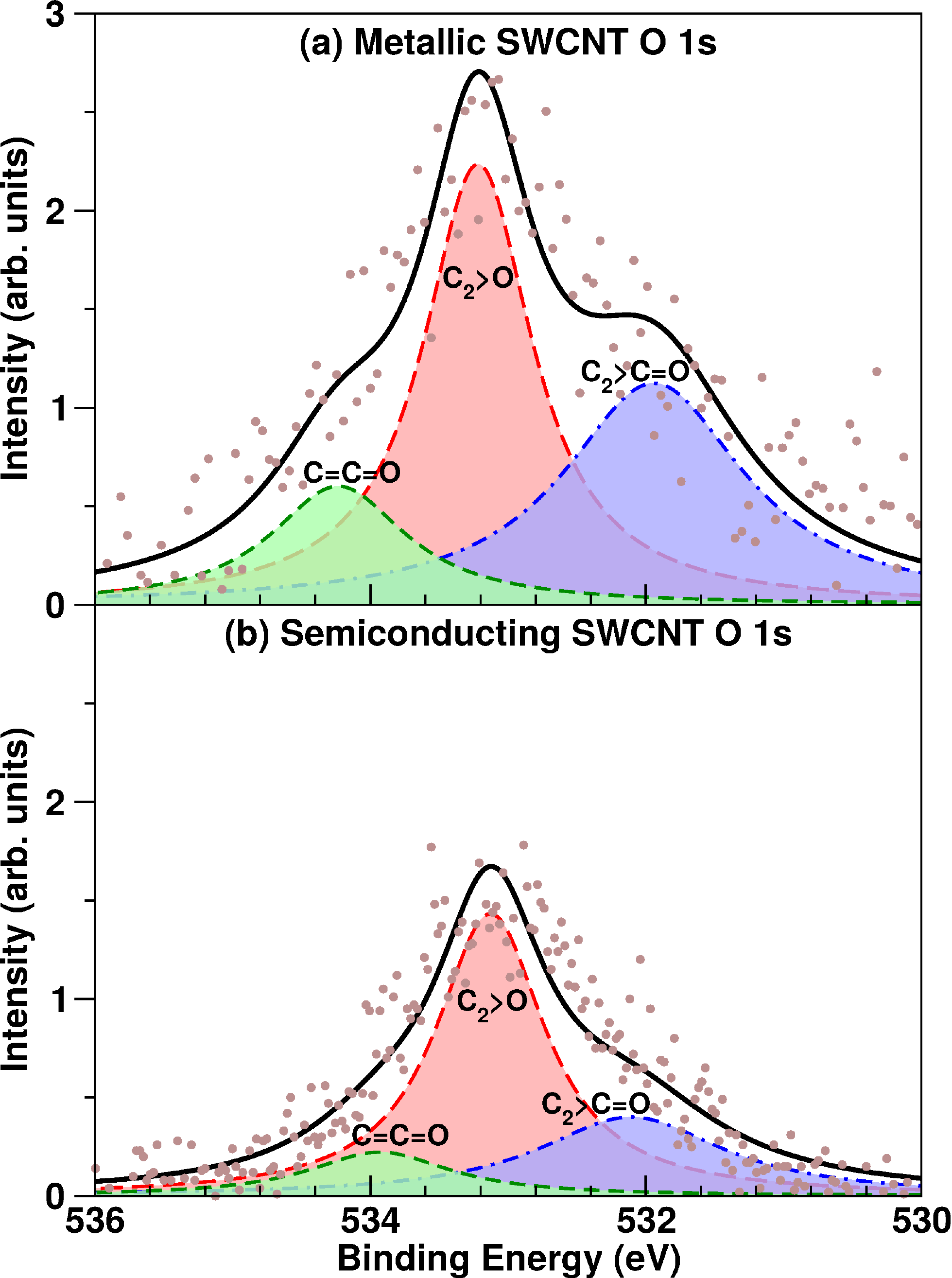}
\caption{Measured O~1s XPS spectra ({\color{brown}{\textbullet}}), Voigtian fit based on DFT eigenvalues (\textbf{---}), and individual components from ketene (\ketene;
\textbf{\color{DarkGreen}{-~-~-}}), epoxy (\epoxy;
\textbf{\color{red}{--~--}}), and carbonyl (\carbonyl;
\textbf{\color{blue}{-~$\cdot$~-~$\cdot$}}) groups for (a) metallic (6,6) and (b) semiconducting (10,0) SWCNTs.}
\label{Fig2}
\noindent{\color{StyleColor}{\rule{\columnwidth}{1.0pt}}}
\end{figure}

Using the O~1s binding energies provided in Table~\ref{Table1},
\begin{table}
\vspace{-16mm}
\noindent{\color{StyleColor}{\rule{\columnwidth}{1.0pt}}}
\vspace{11mm}
\caption{\rm{\bf{Voigtian Fitted Lorentzian Inverse Lifetimes $\boldsymbol\Gamma$ and Intensities \emph{I} to the O~1s XPS Spectra with a Gaussian Broadening $\boldsymbol\sigma$ = 50 meV for O-Containing Species Depicted in Figure~\ref{Fig1} with Areas \emph{A} and Effective Intensities \emph{I}$^{\textbf{\emph{eff}}}$ from Eq.~(\ref{Ieff})}}}\label{table2}
\begin{tabular}{lccccc}
\multicolumn{6}{>{\columncolor[gray]{0.9}}c}{ }\\[-3mm]
\rowcolor[gray]{0.9}Symbol& Species & $\Gamma$ & $I$ & $A$ & $I^{\mathit{eff}}$\\
\rowcolor[gray]{0.9}& 
& (eV) 
& \multicolumn{3}{>{\columncolor[gray]{0.9}}c}{(arbitrary units)}
\\
\multicolumn{6}{c}{Metallic (6,6) SWCNT O~1s:}\\
\ketene        & ketene   & 1.18  & 0.606 & 1.123 &  0.602\\
\epoxy        & epoxy    & 0.95  & 2.259 & 3.371 &  2.235\\
\carbonyl    & carbonyl & 1.59  & 1.128 & 2.817 & 1.124\\

\multicolumn{6}{c}{Semiconducting (10,0) SWCNT O~1s:}\\
\ketene        & ketene    & 1.23   & 0.224  & 0.435 &  0.223\\
\epoxy         & epoxy      & 0.94   & 1.448  & 2.150 &  1.432\\
\carbonyl     & carbonyl  & 1.63   & 0.402  & 1.030 &  0.400\\
\end{tabular}
\noindent{\color{StyleColor}{\rule{\columnwidth}{1.0pt}}}
\end{table}
 we fit the experimental spectra with a sum of Voigtian functions, as shown in Figure~\ref{Fig2}.
The fitting procedure was explained in Section~\ref{FittingProcedure}.  As mentioned previously, the signal-to-noise ratio of the measured XPS spectra is necessarily rather low as our aim is to preferentially oxidize the SWCNT's vacancies.  Physically, we expect the O~1s inverse lifetimes for each O-containing species to be consistent for metallic and semiconducting SWCNTs, and around an electronvolt.

We find a best fit to the measured spectra for metallic and semiconducting SWCNTs exposed to O$_2$ does not include contributions from physisorbed or chemisorbed O$_2$, that is, {\tripletoxygen} or \singletoxygen, respectively.  Of course, this does not rule out the presence of such species, given the difficulty in resolving the signal due to such mobile species.  Further, the epoxy peaks are already well described based solely on the \epoxyp\ binding energies.  However, this does not rule out the presence of \epoxyh\ epoxy species, considering the similarity and accuracy of the calculated binding energies.  For this reason, we do not explicitly specify which site the epoxy group is bound to from here on.

The O~1s inverse lifetimes for each O-containing species are within 60 meV in both the metallic and semiconducting SWCNT samples, as expected.  Overall, the O~1s inverse lifetimes for epoxy and ketene species are about 1 eV, whereas for carbonyl species the O 1s core-hole excitation is shorter lived ($\Gamma \approx 1.6$~eV).

As discussed in Section \ref{Computational Details},  O$_2$ dissociation is facile on SWCNT monovacancies.  This implies, under the experimental conditions employed herein, the intensity of the XPS O~1s signal at saturation is directly correlated with the number of SWCNT vacancies present.  Comparing the intensities of the central peaks in Figure~\ref{Fig2} (a) and (b), we find there are about 60\% more O-containing species on the metallic than the semiconducting SWCNT sample.  This suggests the metallic SWCNT sample is significantly more defective than the semiconducting SWCNT sample.

From the previous discussions, and in order to unravel the relative abundance of various O-containing species in the sample, one should consider the area of the underlying Lorentzian, that is, $A=\pi I \frac{\Gamma}{2}$, from Table~\ref{table2}. In so doing, we remove the experimental error coming from the apparatus $\left(\sigma\to 0\right)$.  For both metallic and semiconducting tubes, we find epoxy species are the most abundant, followed by carbonyl.  Specifically, the area ratios of epoxy to carbonyl and ketene species are about 1.2 and 3.0 for metallic SWCNTs and 2.1 and 4.9 for semiconducting SWCNTs, respectively. Overall, we find the relative abundance of species is more weighted toward epoxy on the semiconducting (10,0) as compared to the metallic (6,6) SWCNTs.  This is consistent with our previous measurements and calculations of the C~1s XPS for the same systems when exposed to NO$_2$\cite{ACSNano}.

\section{CONCLUSIONS}\label{conclusions}
We have demonstrated how by computing the binding energies of various species on SWCNTs, we may disentangle a measured spectrum into its individual components.  This allows one to determine the specific functional groups or bonding environments that are responsible for the measured spectra.  

Here, we used KS O~1s binding energies to unravel the measured XPS O~1s spectra for vacancy oxidation on metallicity-sorted SWCNTs.  We find the XPS O~1s signal is mostly due to three O-containing functional groups: ketene (\ketene), epoxy (\epoxy), and carbonyl (\carbonyl).  The peak intensity for each species on the metallic SWCNT sample is about 60\% greater than that on the semiconducting SWCNT sample.  This suggests a greater abundance of O-containing defect structures on the metallic SWCNT sample.  This agrees with our recent findings for metallicity-sorted SWCNTs exposed to NO$_2$ \cite{ACSNano}.  We find the epoxy, carbonyl, and ketene species have similar lifetimes on both the metallic and semiconducting SWCNT samples.  In both cases we find O$_2$ does not contribute to the decomposed XPS O~1s spectra.  

Using a combined theoretical, computational, and experimental approach, we have shed light on the nature of the O~1s spectra of SWCNTs after oxidizing their vacancies.  This allowed us to disentangle which species are present and which are not, and determine their relative abundance.  Such a combined approach should prove most useful when a clear characterization of the measured system is absent. For example, measurements with low experimental resolution, low signal-to-noise ratio, and/or inhomogeneous samples.  In such cases, applying a computational screening approach to the list of possible species may still allow a robust decomposition of the measured spectra, and the extraction of relevant information otherwise inaccessible via present-day spectroscopic tools.

\titleformat{\section}{\bfseries\sffamily\color{StyleColor}}{\thesection.~}{0pt}{\large$\blacksquare$\normalsize~}
\section*{AUTHOR INFORMATION}
\subsubsection*{Corresponding Authors}
\noindent *Tel.: +34 943 01 8392. E-mail: \href{mailto:duncan.mowbray@gmail.com}{duncan.mowbray@gmail.com}.\\
\noindent *Tel.: +34 943 01 8554. E-mail: \href{mailto:alejandroperezpaz@yahoo.com}{alejandroperezpaz@yahoo.com}.\\ 
\noindent *Tel.: +34 943 01 8292. E-mail: \href{mailto:angel.rubio@ehu.es}{angel.rubio@ehu.es}.

\subsubsection*{Notes}
\noindent The authors declare no competing financial interest.

\section*{ACKNOWLEDGEMENTS}
D.J.M.\ acknowledges funding through the Spanish ``Juan de la Cierva''
program (JCI-2010-08156). A.P.P.\ acknowledges funding through the Proyecto Diputaci\'{o}n Foral de Gipuzkoa (Q4818001B), ``Ayuda  para  la  Especializaci\'{o}n de  Personal  Investigador del  Vicerrectorado  de  Investigaci\'{o}n  de  la  UPV/EHU--2013'', and the Spanish ``Juan de la Cierva-incorporaci\'{o}n'' program (IJCI-2014-20147). P.A.\ was supported by a Marie Curie Intra European Fellowship within the 7th European Community Framework Programme.   A.G.\ thanks the FIRB NANOSOLAR (RBAP11C58Y).  We acknowledge funding by European Projects DYNamo (ERC-2015-AdG-694097), POCAONTAS (FP7-PEOPLE-2012-ITN-316633), MOSTOPHOS (GA no.\ 646259),  and EUSpec (COST Action MP1306); Spanish Grants (FIS2013-46159-C3-1-P) and ``Grupos Consolidados UPV/EHU del Gobierno'' (IT-578-13); and the Air Force Office of Scientific Research (AFOSR) (FA2386-15-1-0006 AOARD 144088).  This work was supported by the Austrian Science Fund through projects FWF P21333-N20, FWF P27769-N20, and FWF NanoBlends I 943-N19 and by EU Proposal no.\ 20105285 for ELETTRA. 


\bibliography{O1sSWCNT}

\providecommand*\mcitethebibliography{\thebibliography}
\csname @ifundefined\endcsname{endmcitethebibliography}
  {\let\endmcitethebibliography\endthebibliography}{}
\begin{mcitethebibliography}{45}
\providecommand*\natexlab[1]{#1}
\providecommand*\mciteSetBstSublistMode[1]{}
\providecommand*\mciteSetBstMaxWidthForm[2]{}
\providecommand*\mciteBstWouldAddEndPuncttrue
  {\def\EndOfBibitem{\unskip.}}
\providecommand*\mciteBstWouldAddEndPunctfalse
  {\let\EndOfBibitem\relax}
\providecommand*\mciteSetBstMidEndSepPunct[3]{}
\providecommand*\mciteSetBstSublistLabelBeginEnd[3]{}
\providecommand*\EndOfBibitem{}
\mciteSetBstSublistMode{f}
\mciteSetBstMaxWidthForm{subitem}{(\alph{mcitesubitemcount})}
\mciteSetBstSublistLabelBeginEnd
  {\mcitemaxwidthsubitemform\space}
  {\relax}
  {\relax}

\bibitem[Iijima(1991)]{Iijima}
Iijima,~S. Helical Microtubules of Graphitic Carbon.
  \href{http://dx.doi.org/dx.doi.org/10.1038/354056a0}{\emph{Nature}}
  \textbf{1991}, \emph{354}, 56--68\relax
\mciteBstWouldAddEndPuncttrue
\mciteSetBstMidEndSepPunct{\mcitedefaultmidpunct}
{\mcitedefaultendpunct}{\mcitedefaultseppunct}\relax
\EndOfBibitem
\bibitem[Harris(1999)]{Harris}
Harris,~P. J.~F. \emph{Carbon Nanotubes and Related Structures: New Materials
  for the Twenty-first Century}; Cambridge University Press: Cambridge,
  1999\relax
\mciteBstWouldAddEndPuncttrue
\mciteSetBstMidEndSepPunct{\mcitedefaultmidpunct}
{\mcitedefaultendpunct}{\mcitedefaultseppunct}\relax
\EndOfBibitem
\bibitem[Dresselhaus \textit{et~al.}(2001)Dresselhaus, Dresselhaus, and
  Avouris]{Dresselhaus}
Dresselhaus,~M.~S., Dresselhaus,~G., Avouris,~P., Eds. \emph{Carbon Nanotubes:
  Synthesis, Structure, Properties, and Applications}; Springer: Berlin,
  2001\relax
\mciteBstWouldAddEndPuncttrue
\mciteSetBstMidEndSepPunct{\mcitedefaultmidpunct}
{\mcitedefaultendpunct}{\mcitedefaultseppunct}\relax
\EndOfBibitem
\bibitem[Garc\'{\i}a-Lastra \textit{et~al.}(2008)Garc\'{\i}a-Lastra, Thygesen,
  Strange, and Rubio]{JuanmaPRL}
Garc\'{\i}a-Lastra,~J.~M.; Thygesen,~K.~S.; Strange,~M.; Rubio,~{\'{A}}.
  Conductance of Sidewall-Functionalized Carbon Nanotubes: Universal Dependence
  on Adsorption Sites.
  \href{http://dx.doi.org/10.1103/PhysRevLett.101.236806}{\emph{Phys. Rev.
  Lett.}} \textbf{2008}, \emph{101}, 236806\relax
\mciteBstWouldAddEndPuncttrue
\mciteSetBstMidEndSepPunct{\mcitedefaultmidpunct}
{\mcitedefaultendpunct}{\mcitedefaultseppunct}\relax
\EndOfBibitem
\bibitem[Srivastava \textit{et~al.}(2014)Srivastava, Susi, Borghei, and
  Kari]{TomaOdissociationSWNT}
Srivastava,~D.; Susi,~T.; Borghei,~M.; Kari,~L. Dissociation of Oxygen on
  Pristine and Nitrogen-Doped Carbon Nanotubes: A Spin-Polarized Density
  Functional Study. \href{http://dx.doi.org/10.1039/C3RA47784C}{\emph{RSC
  Adv.}} \textbf{2014}, \emph{4}, 15225--15235\relax
\mciteBstWouldAddEndPuncttrue
\mciteSetBstMidEndSepPunct{\mcitedefaultmidpunct}
{\mcitedefaultendpunct}{\mcitedefaultseppunct}\relax
\EndOfBibitem
\bibitem[Ni \textit{et~al.}(2012)Ni, Li, and Yang]{OdissociationSWNT}
Ni,~S.; Li,~Z.; Yang,~J. Oxygen Molecule Dissociation on Carbon Nanostructures
  with Different Types of Nitrogen Doping.
  \href{http://dx.doi.org/10.1039/C1NR11086A}{\emph{Nanoscale}} \textbf{2012},
  \emph{4}, 1184--1189\relax
\mciteBstWouldAddEndPuncttrue
\mciteSetBstMidEndSepPunct{\mcitedefaultmidpunct}
{\mcitedefaultendpunct}{\mcitedefaultseppunct}\relax
\EndOfBibitem
\bibitem[G\"{u}rel \textit{et~al.}(2014)G\"{u}rel, \"{O}z\c{c}elik, and
  Ciraci]{O2dissociationGrapheneMV}
G\"{u}rel,~H.~H.; \"{O}z\c{c}elik,~V.~O.; Ciraci,~S. Dissociative Adsorption of
  Molecules on Graphene and Silicene.
  \href{http://dx.doi.org/10.1021/jp509260c}{\emph{J. Phys. Chem. C}}
  \textbf{2014}, \emph{118}, 27574--27582\relax
\mciteBstWouldAddEndPuncttrue
\mciteSetBstMidEndSepPunct{\mcitedefaultmidpunct}
{\mcitedefaultendpunct}{\mcitedefaultseppunct}\relax
\EndOfBibitem
\bibitem[Garc\'{\i}a-Lastra \textit{et~al.}(2010)Garc\'{\i}a-Lastra, Mowbray,
  Thygesen, Rubio, and Jacobsen]{GarciaLastra10PRB}
Garc\'{\i}a-Lastra,~J.~M.; Mowbray,~D.~J.; Thygesen,~K.~S.; Rubio,~A.;
  Jacobsen,~K.~W. Modeling Nanoscale Gas Sensors Under Realistic Conditions:
  Computational Screening of Metal-Doped Carbon Nanotubes.
  \href{http://dx.doi.org/10.1103/PhysRevB.81.245429}{\emph{Phys. Rev. B:
  Condens. Matter Mater. Phys.}} \textbf{2010}, \emph{81}, 245429\relax
\mciteBstWouldAddEndPuncttrue
\mciteSetBstMidEndSepPunct{\mcitedefaultmidpunct}
{\mcitedefaultendpunct}{\mcitedefaultseppunct}\relax
\EndOfBibitem
\bibitem[Bondavalli \textit{et~al.}(2009)Bondavalli, Legagneux, and
  Pribat]{Bondavalli09SABC}
Bondavalli,~P.; Legagneux,~P.; Pribat,~D. Carbon Nanotubes Based Transistors as
  Gas Sensors: State of the Art and Critical Review.
  \href{http://dx.doi.org/10.1016/j.snb.2009.04.025}{\emph{Sens. Act. B -
  Chemical}} \textbf{2009}, \emph{140}, 304\relax
\mciteBstWouldAddEndPuncttrue
\mciteSetBstMidEndSepPunct{\mcitedefaultmidpunct}
{\mcitedefaultendpunct}{\mcitedefaultseppunct}\relax
\EndOfBibitem
\bibitem[Jhi \textit{et~al.}(2000)Jhi, Louie, and Cohen]{Jhi00PRL}
Jhi,~S.~H.; Louie,~S.~G.; Cohen,~M.~L. Electronic Properties of Oxidized Carbon
  Nanotubes. \href{http://dx.doi.org/10.1103/PhysRevLett.85.1710}{\emph{Phys.
  Rev. Lett.}} \textbf{2000}, \emph{85}, 1710\relax
\mciteBstWouldAddEndPuncttrue
\mciteSetBstMidEndSepPunct{\mcitedefaultmidpunct}
{\mcitedefaultendpunct}{\mcitedefaultseppunct}\relax
\EndOfBibitem
\bibitem[Collins \textit{et~al.}(2000)Collins, Bradley, Ishigami, and
  Zettl]{Collins00S}
Collins,~P.~G.; Bradley,~K.; Ishigami,~M.; Zettl,~A. Extreme Oxygen Sensitivity
  of Electronic Properties of Carbon Nanotubes.
  \href{http://dx.doi.org/10.1126/science.287.5459.1801}{\emph{Science}}
  \textbf{2000}, \emph{287}, 1801\relax
\mciteBstWouldAddEndPuncttrue
\mciteSetBstMidEndSepPunct{\mcitedefaultmidpunct}
{\mcitedefaultendpunct}{\mcitedefaultseppunct}\relax
\EndOfBibitem
\bibitem[Ulbricht \textit{et~al.}(2002)Ulbricht, Moos, and
  Hertel]{Ulbricht02PRB}
Ulbricht,~H.; Moos,~G.; Hertel,~T. Physisorption of Molecular Oxygen on
  Single-Wall Carbon Nanotube Bundles and Graphite.
  \href{http://dx.doi.org/10.1103/PhysRevB.66.075404}{\emph{Phys. Rev. B:
  Condens. Matter Mater. Phys.}} \textbf{2002}, \emph{66}, 075404\relax
\mciteBstWouldAddEndPuncttrue
\mciteSetBstMidEndSepPunct{\mcitedefaultmidpunct}
{\mcitedefaultendpunct}{\mcitedefaultseppunct}\relax
\EndOfBibitem
\bibitem[Goldoni \textit{et~al.}(2003)Goldoni, Larciprete, Petaccia, and
  Lizzit]{Goldoni03JACS}
Goldoni,~A.; Larciprete,~R.; Petaccia,~L.; Lizzit,~S. Single-Wall Carbon
  Nanotube Interaction with Gases: Sample Contaminants and Environmental
  Monitoring. \href{http://dx.doi.org/10.1021/ja034898e}{\emph{J. Am. Chem.
  Soc.}} \textbf{2003}, \emph{125}, 11329\relax
\mciteBstWouldAddEndPuncttrue
\mciteSetBstMidEndSepPunct{\mcitedefaultmidpunct}
{\mcitedefaultendpunct}{\mcitedefaultseppunct}\relax
\EndOfBibitem
\bibitem[Dag \textit{et~al.}(2003)Dag, Goulseren, Yildirim, and
  Ciraci]{Dag2003}
Dag,~S.; Goulseren,~O.; Yildirim,~T.; Ciraci,~S. Oxygenation of Carbon
  Nanotubes: Atomic Structure, Energetics, and Electronic Structure.
  \href{http://dx.doi.org/10.1103/PhysRevB.67.165424}{\emph{Phys. Rev. B:
  Condens. Matter Mater. Phys.}} \textbf{2003}, \emph{67}, 165424\relax
\mciteBstWouldAddEndPuncttrue
\mciteSetBstMidEndSepPunct{\mcitedefaultmidpunct}
{\mcitedefaultendpunct}{\mcitedefaultseppunct}\relax
\EndOfBibitem
\bibitem[Grujicic \textit{et~al.}(2003)Grujicic, Cao, Rao, Tritt, and
  Nayak]{Grujicic03ASS}
Grujicic,~M.; Cao,~G.; Rao,~A.; Tritt,~T.; Nayak,~S. UV-Light Enhanced
  Oxidation of Carbon Nanotubes.
  \href{http://dx.doi.org/10.1016/S0169-4332(03)00361-1}{\emph{Appl. Surf.
  Sci.}} \textbf{2003}, \emph{214}, 289\relax
\mciteBstWouldAddEndPuncttrue
\mciteSetBstMidEndSepPunct{\mcitedefaultmidpunct}
{\mcitedefaultendpunct}{\mcitedefaultseppunct}\relax
\EndOfBibitem
\bibitem[Giannozzi \textit{et~al.}(2003)Giannozzi, Car, and
  Scoles]{Giannozzi03JCP}
Giannozzi,~P.; Car,~R.; Scoles,~G. Oxygen Adsorption on Graphite and Nanotubes.
  \href{http://dx.doi.org/10.1063/1.1536636}{\emph{J. Chem. Phys.}}
  \textbf{2003}, \emph{118}, 1003\relax
\mciteBstWouldAddEndPuncttrue
\mciteSetBstMidEndSepPunct{\mcitedefaultmidpunct}
{\mcitedefaultendpunct}{\mcitedefaultseppunct}\relax
\EndOfBibitem
\bibitem[Datsyuk \textit{et~al.}(2008)Datsyuk, Kalyva, Papagelis, Parthenios,
  Tasis, Siokou, Kallitsis, and Galiotis]{Datsyuk08C}
Datsyuk,~V.; Kalyva,~M.; Papagelis,~K.; Parthenios,~J.; Tasis,~D.; Siokou,~A.;
  Kallitsis,~I.; Galiotis,~C. Chemical Oxidation of Multiwalled Carbon
  Nanotubes.
  \href{http://dx.doi.org/10.1016/j.carbon.2008.02.012}{\emph{Carbon}}
  \textbf{2008}, \emph{46}, 833--840\relax
\mciteBstWouldAddEndPuncttrue
\mciteSetBstMidEndSepPunct{\mcitedefaultmidpunct}
{\mcitedefaultendpunct}{\mcitedefaultseppunct}\relax
\EndOfBibitem
\bibitem[Mowbray \textit{et~al.}(2009)Mowbray, Morgan, and
  Thygesen]{Mowbray09PRB}
Mowbray,~D.~J.; Morgan,~C.; Thygesen,~K.~S. Influence of ${\text{O}}_{2}$ and
  ${\text{N}}_{2}$ on the Conductivity of Carbon Nanotube Networks.
  \href{http://dx.doi.org/10.1103/PhysRevB.79.195431}{\emph{Phys. Rev. B:
  Condens. Matter Mater. Phys.}} \textbf{2009}, \emph{79}, 195431\relax
\mciteBstWouldAddEndPuncttrue
\mciteSetBstMidEndSepPunct{\mcitedefaultmidpunct}
{\mcitedefaultendpunct}{\mcitedefaultseppunct}\relax
\EndOfBibitem
\bibitem[Kroes \textit{et~al.}(2016)Kroes, Pietrucci, Chikkadi, Roman, Hierold,
  and Andreoni]{WandaAndreoni}
Kroes,~J. M.~H.; Pietrucci,~F.; Chikkadi,~K.; Roman,~C.; Hierold,~C.;
  Andreoni,~W. The Response of Single-Walled Carbon Nanotubes to NO$_2$ and the
  Search for a Long-Living Adsorbed Species.
  \href{http://dx.doi.org/10.1063/1.4940422}{\emph{Appl. Phys. Lett.}}
  \textbf{2016}, \emph{108}, 033111\relax
\mciteBstWouldAddEndPuncttrue
\mciteSetBstMidEndSepPunct{\mcitedefaultmidpunct}
{\mcitedefaultendpunct}{\mcitedefaultseppunct}\relax
\EndOfBibitem
\bibitem[Ruiz-Soria \textit{et~al.}(2014)Ruiz-Soria, P\'{e}rez~Paz, Sauer,
  Mowbray, Lacovig, Dalmiglio, Lizzit, Yanagi, Rubio, Goldoni, Ayala, and
  Pichler]{ACSNano}
Ruiz-Soria,~G. \textit{et~al.}  Revealing the Adsorption Mechanisms of
  Nitroxides on Ultra-Pure, Metallicity-Sorted Carbon Nanotubes.
  \href{http://dx.doi.org/10.1021/nn405114z}{\emph{ACS Nano}} \textbf{2014},
  \emph{8}, 1375--1383\relax
\mciteBstWouldAddEndPuncttrue
\mciteSetBstMidEndSepPunct{\mcitedefaultmidpunct}
{\mcitedefaultendpunct}{\mcitedefaultseppunct}\relax
\EndOfBibitem
\bibitem[Van~den Bossche \textit{et~al.}(2014)Van~den Bossche, Martin,
  Gustafson, Hakanoglu, Weaver, Lundgren, and Gr{\"{o}}nbeck]{HybridCLSJPC2014}
Van~den Bossche,~M.; Martin,~N.~M.; Gustafson,~J.; Hakanoglu,~C.;
  Weaver,~J.~F.; Lundgren,~E.; Gr{\"{o}}nbeck,~H. Effects of Non-Local Exchange
  on Core Level Shifts for Gas-Phase and Adsorbed Molecules.
  \href{http://dx.doi.org/http://dx.doi.org/10.1063/1.4889919}{\emph{J. Chem.
  Phys.}} \textbf{2014}, \emph{141}, 034706\relax
\mciteBstWouldAddEndPuncttrue
\mciteSetBstMidEndSepPunct{\mcitedefaultmidpunct}
{\mcitedefaultendpunct}{\mcitedefaultseppunct}\relax
\EndOfBibitem
\bibitem[Binder \textit{et~al.}(2012)Binder, Broqvist, Komsa, and
  Pasquarello]{HybridCLSPRB2012}
Binder,~J.~F.; Broqvist,~P.; Komsa,~H.-P.; Pasquarello,~A. Germanium Core-Level
  Shifts at Ge/GeO${}_{2}$ Interfaces Through Hybrid Functionals.
  \href{http://dx.doi.org/10.1103/PhysRevB.85.245305}{\emph{Phys. Rev. B:
  Condens. Matter Mater. Phys.}} \textbf{2012}, \emph{85}, 245305\relax
\mciteBstWouldAddEndPuncttrue
\mciteSetBstMidEndSepPunct{\mcitedefaultmidpunct}
{\mcitedefaultendpunct}{\mcitedefaultseppunct}\relax
\EndOfBibitem
\bibitem[Paz-Borb{\'{o}}n \textit{et~al.}(2012)Paz-Borb{\'{o}}n, Hellman, and
  Gr{\"{o}}nbeck]{PBEvsPBE0CLSJPCC2012}
Paz-Borb{\'{o}}n,~L.~O.; Hellman,~A.; Gr{\"{o}}nbeck,~H. Simulated
  Photoemission Spectra of Hydroxylated MgO(100) at Elevated Temperatures.
  \href{http://dx.doi.org/10.1021/jp209336q}{\emph{J. Phys. Chem. C}}
  \textbf{2012}, \emph{116}, 3545--3551\relax
\mciteBstWouldAddEndPuncttrue
\mciteSetBstMidEndSepPunct{\mcitedefaultmidpunct}
{\mcitedefaultendpunct}{\mcitedefaultseppunct}\relax
\EndOfBibitem
\bibitem[Ping \textit{et~al.}(2015)Ping, Galli, and William
  A.~Goddard]{PBEvsPBEUJPCC2015}
Ping,~Y.; Galli,~G.; William A.~Goddard,~I. Electronic Structure of IrO$_{2}$:
  The Role of the Metal d Orbitals.
  \href{http://dx.doi.org/10.1021/acs.jpcc.5b00861}{\emph{J. Phys. Chem. C}}
  \textbf{2015}, \emph{119}, 11570--11577\relax
\mciteBstWouldAddEndPuncttrue
\mciteSetBstMidEndSepPunct{\mcitedefaultmidpunct}
{\mcitedefaultendpunct}{\mcitedefaultseppunct}\relax
\EndOfBibitem
\bibitem[Cabellos \textit{et~al.}(2012)Cabellos, Mowbray, Goiri, El-Sayed,
  Floreano, de~Oteyza, Rogero, Ortega, and Rubio]{Cabellos12JPCC}
Cabellos,~J.~L.; Mowbray,~D.~J.; Goiri,~E.; El-Sayed,~A.; Floreano,~L.;
  de~Oteyza,~D.~G.; Rogero,~C.; Ortega,~J.~E.; Rubio,~A. Understanding Charge
  Transfer in Donor--Acceptor/Metal Systems: A Combined Theoretical and
  Experimental Study. \href{http://dx.doi.org/10.1021/jp3004213}{\emph{J. Phys.
  Chem. C}} \textbf{2012}, \emph{116}, 17991--18001\relax
\mciteBstWouldAddEndPuncttrue
\mciteSetBstMidEndSepPunct{\mcitedefaultmidpunct}
{\mcitedefaultendpunct}{\mcitedefaultseppunct}\relax
\EndOfBibitem
\bibitem[Haubner \textit{et~al.}(2010)Haubner, Murawski, Olk, Eng, Ziegler,
  Adolphi, and Jaehne]{kinga}
Haubner,~K.; Murawski,~J.; Olk,~P.; Eng,~L.~M.; Ziegler,~C.; Adolphi,~B.;
  Jaehne,~E. The Route to Functional Graphene Oxide.
  \href{http://dx.doi.org/10.1002/cphc.201000132}{\emph{ChemPhysChem}}
  \textbf{2010}, \emph{11}, 2131--2139\relax
\mciteBstWouldAddEndPuncttrue
\mciteSetBstMidEndSepPunct{\mcitedefaultmidpunct}
{\mcitedefaultendpunct}{\mcitedefaultseppunct}\relax
\EndOfBibitem
\bibitem[Rozada \textit{et~al.}(2013)Rozada, Paredes, Villar-Rodil,
  Mart\'{i}nez-Alonso, and Tasc\'{o}n]{rozada}
Rozada,~R.; Paredes,~J.~I.; Villar-Rodil,~S.; Mart\'{i}nez-Alonso,~A.;
  Tasc\'{o}n,~J. M.~D. Towards Full Repair of Defects in Reduced Graphene Oxide
  Films by Two-Step Graphitization.
  \href{http://dx.doi.org/10.1007/s12274-013-0298-6}{\emph{Nano Res.}}
  \textbf{2013}, \emph{6}, 216--233\relax
\mciteBstWouldAddEndPuncttrue
\mciteSetBstMidEndSepPunct{\mcitedefaultmidpunct}
{\mcitedefaultendpunct}{\mcitedefaultseppunct}\relax
\EndOfBibitem
\bibitem[Vinogradov \textit{et~al.}(2011)Vinogradov, Schulte, Ng, Mikkelsen,
  Lundgren, M\r{a}rtensson, and Preobrajenski]{Vinogradov}
Vinogradov,~N.~A.; Schulte,~K.; Ng,~M.~L.; Mikkelsen,~A.; Lundgren,~E.;
  M\r{a}rtensson,~N.; Preobrajenski,~A.~B. Impact of Atomic Oxygen on the
  Structure of Graphene Formed on {Ir}(111) and {Pt}(111).
  \href{http://dx.doi.org/10.1021/jp111962k}{\emph{J. Phys. Chem. C}}
  \textbf{2011}, \emph{115}, 9568--9577\relax
\mciteBstWouldAddEndPuncttrue
\mciteSetBstMidEndSepPunct{\mcitedefaultmidpunct}
{\mcitedefaultendpunct}{\mcitedefaultseppunct}\relax
\EndOfBibitem
\bibitem[Xing \textit{et~al.}(2005)Xing, Li, Chusuei, and
  Hull]{LangmuirO1s2005}
Xing,~Y.; Li,~L.; Chusuei,~C.~C.; Hull,~R.~V. Sonochemical Oxidation of
  Multiwalled Carbon Nanotubes.
  \href{http://dx.doi.org/10.1021/la047268e}{\emph{Langmuir}} \textbf{2005},
  \emph{21}, 4185--4190\relax
\mciteBstWouldAddEndPuncttrue
\mciteSetBstMidEndSepPunct{\mcitedefaultmidpunct}
{\mcitedefaultendpunct}{\mcitedefaultseppunct}\relax
\EndOfBibitem
\bibitem[Mart\'{\i}nez \textit{et~al.}(2003)Mart\'{\i}nez, Callejas, Benito,
  Cochet, Seeger, Ans\'{o}n, Schreiber, Gordon, Marhic, Chauvet, Fierro, and
  Maser]{Martinez03}
Mart\'{\i}nez,~M. \textit{et~al.}  Sensitivity of Single Wall Carbon Nanotubes
  to Oxidative Processing: Structural Modification, Intercalation and
  Functionalisation.
  \href{http://dx.doi.org/10.1016/S0008-6223(03)00250-1}{\emph{Carbon}}
  \textbf{2003}, \emph{41}, 2247--2256\relax
\mciteBstWouldAddEndPuncttrue
\mciteSetBstMidEndSepPunct{\mcitedefaultmidpunct}
{\mcitedefaultendpunct}{\mcitedefaultseppunct}\relax
\EndOfBibitem
\bibitem[Markiewicz \textit{et~al.}(2014)Markiewicz, Wilczewska, Chernyaeva,
  and Winkler]{Markiewicz_PCCP2014}
Markiewicz,~K.~H.; Wilczewska,~A.~Z.; Chernyaeva,~O.; Winkler,~K. Ring-Opening
  Reactions of Epoxidized SWCNT with Nucleophilic Agents: A Convenient Way for
  Sidewall Functionalization.
  \href{http://dx.doi.org/10.1039/C4NJ00148F}{\emph{New J. Chem.}}
  \textbf{2014}, \emph{38}, 2670--2678\relax
\mciteBstWouldAddEndPuncttrue
\mciteSetBstMidEndSepPunct{\mcitedefaultmidpunct}
{\mcitedefaultendpunct}{\mcitedefaultseppunct}\relax
\EndOfBibitem
\bibitem[Abrami \textit{et~al.}(1995)Abrami, Barnaba, Battistello, Bianco,
  Brena, Cautero, Chen, Cocco, Comelli, Contrino, Debona, Difonzo, Fava,
  Finetti, Furlan, Galimberti, Gambitta, Giuressi, Godnig, Jark, Lizzit,
  Mazzolini, Melpignano, Olivi, Paolucci, Pugliese, Qian, Rosei, Sandrin,
  Savoia, Sergo, Sostero, Tommasini, Tudor, Vivoda, Wei, and
  Zanini]{Abrami95RSI}
Abrami,~A. \textit{et~al.}  Super {ESCA} - First Beamline Operating At
  {ELETTRA}. \href{http://dx.doi.org/10.1063/1.1145862}{\emph{Rev. Sci.
  Instrum.}} \textbf{1995}, \emph{66}, 1618\relax
\mciteBstWouldAddEndPuncttrue
\mciteSetBstMidEndSepPunct{\mcitedefaultmidpunct}
{\mcitedefaultendpunct}{\mcitedefaultseppunct}\relax
\EndOfBibitem
\bibitem[Ayala \textit{et~al.}(2009)Ayala, Miyata, De~Blauwe, Shiozawa, Feng,
  Yanagi, Kramberger, Silva, Follath, Kataura, and Pichler]{Ayala09PRB}
Ayala,~P. \textit{et~al.}  Disentanglement of the Electronic Properties of
  Metallicity-Selected Single-Walled Carbon Nanotubes.
  \href{http://dx.doi.org/10.1103/PhysRevB.80.205427}{\emph{Phys. Rev. B:
  Condens. Matter Mater. Phys.}} \textbf{2009}, \emph{80}, 205427\relax
\mciteBstWouldAddEndPuncttrue
\mciteSetBstMidEndSepPunct{\mcitedefaultmidpunct}
{\mcitedefaultendpunct}{\mcitedefaultseppunct}\relax
\EndOfBibitem
\bibitem[Kramberger \textit{et~al.}(2007)Kramberger, Rauf, Shiozawa, Knupfer,
  Buchner, Pichler, Batchelor, and Kataura]{Kramberger07PRB}
Kramberger,~C.; Rauf,~H.; Shiozawa,~H.; Knupfer,~M.; Buchner,~B.; Pichler,~T.;
  Batchelor,~D.; Kataura,~H. Unraveling van Hove Singularities in X-Ray
  Absorption Response of Single-Wall Carbon Nanotubes.
  \href{http://dx.doi.org/10.1103/PhysRevB.75.235437}{\emph{Phys. Rev. B:
  Condens. Matter Mater. Phys..}} \textbf{2007}, \emph{75}, 235437\relax
\mciteBstWouldAddEndPuncttrue
\mciteSetBstMidEndSepPunct{\mcitedefaultmidpunct}
{\mcitedefaultendpunct}{\mcitedefaultseppunct}\relax
\EndOfBibitem
\bibitem[Mortensen \textit{et~al.}(2005)Mortensen, Hansen, and
  Jacobsen]{Mortensen05PRB}
Mortensen,~J.~J.; Hansen,~L.~B.; Jacobsen,~K.~W. Real-Space Grid Implementation
  of the Projector Augmented Wave Method.
  \href{http://dx.doi.org/10.1103/PhysRevB.71.035109}{\emph{Phys. Rev. B:
  Condens. Matter Mater. Phys.}} \textbf{2005}, \emph{71}, 035109\relax
\mciteBstWouldAddEndPuncttrue
\mciteSetBstMidEndSepPunct{\mcitedefaultmidpunct}
{\mcitedefaultendpunct}{\mcitedefaultseppunct}\relax
\EndOfBibitem
\bibitem[Enkovaara \textit{et~al.}(2010)Enkovaara, Rostgaard, Mortensen, Chen,
  Du{\/l}ak, Ferrighi, Gavnholt, Glinsvad, Haikola, Hansen, Kristoffersen,
  Kuisma, Larsen, Lehtovaara, Ljungberg, Lopez-Acevedo, Moses, Ojanen, Olsen,
  Petzold, Romero, Stausholm-M{\o}ller, Strange, Tritsaris, Vanin, Walter,
  Hammer, H\"{a}kkinen, Madsen, Nieminen, N{\o}rskov, Puska, Rantala,
  Schi{\o}tz, Thygesen, and Jacobsen]{Enkovaara10JPCM}
Enkovaara,~J. \textit{et~al.}  Electronic Structure Calculations with {GPAW}: A
  Real-Space Implementation of the Projector Augmented-Wave Method.
  \href{http://dx.doi.org/10.1088/0953-8984/22/25/253202}{\emph{J. Phys.:
  Condens. Matter}} \textbf{2010}, \emph{22}, 253202\relax
\mciteBstWouldAddEndPuncttrue
\mciteSetBstMidEndSepPunct{\mcitedefaultmidpunct}
{\mcitedefaultendpunct}{\mcitedefaultseppunct}\relax
\EndOfBibitem
\bibitem[Bahn and Jacobsen(2002)Bahn, and Jacobsen]{Bahn02CSE}
Bahn,~S.~R.; Jacobsen,~K.~W. An Object-Oriented Scripting Interface to a Legacy
  Electronic Structure Code.
  \href{http://dx.doi.org/10.1109/5992.998641}{\emph{Comput. Sci. Eng.}}
  \textbf{2002}, \emph{4}, 56\relax
\mciteBstWouldAddEndPuncttrue
\mciteSetBstMidEndSepPunct{\mcitedefaultmidpunct}
{\mcitedefaultendpunct}{\mcitedefaultseppunct}\relax
\EndOfBibitem
\bibitem[Perdew and Zunger(1981)Perdew, and Zunger]{Perdew81PRB}
Perdew,~J.~P.; Zunger,~A. Self-Interaction Correction to Density-Functional
  Approximations for Many-Electron Systems.
  \href{http://dx.doi.org/10.1103/PhysRevB.23.5048}{\emph{Phys. Rev. B:
  Condens. Matter Mater. Phys.}} \textbf{1981}, \emph{23}, 5048\relax
\mciteBstWouldAddEndPuncttrue
\mciteSetBstMidEndSepPunct{\mcitedefaultmidpunct}
{\mcitedefaultendpunct}{\mcitedefaultseppunct}\relax
\EndOfBibitem
\bibitem[Susi \textit{et~al.}(2015)Susi, Mowbray, Ljungberg, and
  Ayala]{TomaPRB2015}
Susi,~T.; Mowbray,~D.~J.; Ljungberg,~M.~P.; Ayala,~P. Calculation of the
  Graphene C $1s$ Core Level Binding Energy.
  \href{http://dx.doi.org/10.1103/PhysRevB.91.081401}{\emph{Phys. Rev. B:
  Condens. Matter Mater. Phys.}} \textbf{2015}, \emph{91}, 081401\relax
\mciteBstWouldAddEndPuncttrue
\mciteSetBstMidEndSepPunct{\mcitedefaultmidpunct}
{\mcitedefaultendpunct}{\mcitedefaultseppunct}\relax
\EndOfBibitem
\bibitem[Zhang and Yang(1998)Zhang, and Yang]{revPBE}
Zhang,~Y.; Yang,~W. Comment on ``Generalized Gradient Approximation Made
  Simple''. \href{http://dx.doi.org/10.1103/PhysRevLett.80.890}{\emph{Phys.
  Rev. Lett.}} \textbf{1998}, \emph{80}, 890--890\relax
\mciteBstWouldAddEndPuncttrue
\mciteSetBstMidEndSepPunct{\mcitedefaultmidpunct}
{\mcitedefaultendpunct}{\mcitedefaultseppunct}\relax
\EndOfBibitem
\bibitem[Sorensen \textit{et~al.}(2008)Sorensen, B{\o}rve, Feifel, de~Fanis,
  and Ueda]{Sorensen}
Sorensen,~S.~L.; B{\o}rve,~K.~J.; Feifel,~R.; de~Fanis,~A.; Ueda,~K. The {O~1s}
  Photoelectron Spectrum of Molecular Oxygen Revisited.
  \href{http://dx.doi.org/10.1088/0953-4075/41/9/095101}{\emph{J. Phys. B: At.
  Mol. Opt. Phys.}} \textbf{2008}, \emph{41}, 095101\relax
\mciteBstWouldAddEndPuncttrue
\mciteSetBstMidEndSepPunct{\mcitedefaultmidpunct}
{\mcitedefaultendpunct}{\mcitedefaultseppunct}\relax
\EndOfBibitem
\bibitem[Henkelman \textit{et~al.}(2006)Henkelman, Arnaldsson, and
  J\'{o}nsson]{Henkelman2006354}
Henkelman,~G.; Arnaldsson,~A.; J\'{o}nsson,~H. A Fast and Robust Algorithm for
  {Bader} Decomposition of Charge Density.
  \href{http://dx.doi.org/10.1016/j.commatsci.2005.04.010}{\emph{Comput. Mater.
  Sci.}} \textbf{2006}, \emph{36}, 354 -- 360\relax
\mciteBstWouldAddEndPuncttrue
\mciteSetBstMidEndSepPunct{\mcitedefaultmidpunct}
{\mcitedefaultendpunct}{\mcitedefaultseppunct}\relax
\EndOfBibitem
\bibitem[Wu \textit{et~al.}(2015)Wu, Wen, Schlogl, and Su]{PCCPO1sSWNTs2015}
Wu,~S.; Wen,~G.; Schlogl,~R.; Su,~D.~S. Carbon Nanotubes Oxidized by a Green
  Method as Efficient Metal-Free Catalysts for Nitroarene Reduction.
  \href{http://dx.doi.org/10.1039/C4CP04658G}{\emph{Phys. Chem. Chem. Phys.}}
  \textbf{2015}, \emph{17}, 1567--1571\relax
\mciteBstWouldAddEndPuncttrue
\mciteSetBstMidEndSepPunct{\mcitedefaultmidpunct}
{\mcitedefaultendpunct}{\mcitedefaultseppunct}\relax
\EndOfBibitem
\bibitem[El-Sayed \textit{et~al.}(2013)El-Sayed, Borghetti, Goiri, Rogero,
  Floreano, Lovat, Mowbray, Cabellos, Wakayama, Rubio, Ortega, and
  de~Oteyza]{MowbrayACSNano}
El-Sayed,~A. \textit{et~al.}  Understanding Energy Level Alignment in
  Donor-Acceptor/Metal Interfaces from Core-Level Shifts.
  \href{http://dx.doi.org/10.1021/nn4020888}{\emph{ACS Nano}} \textbf{2013},
  \emph{7}, 6914--6920\relax
\mciteBstWouldAddEndPuncttrue
\mciteSetBstMidEndSepPunct{\mcitedefaultmidpunct}
{\mcitedefaultendpunct}{\mcitedefaultseppunct}\relax
\EndOfBibitem
\end{mcitethebibliography}


\end{document}